\documentclass[12pt]{article}
\usepackage{epsfig,amssymb,amsmath,psfrag,subfigure,rotate,color,wasysym,graphicx}

\allowdisplaybreaks
\newcommand{\insertfig}[2]{\includegraphics[width=#1cm]{#2}}

\DeclareMathOperator*{\SumInt}{%
\mathchoice%
  {\ooalign{$\displaystyle\sum$\cr\hidewidth$\displaystyle\int$\hidewidth\cr}}
  {\ooalign{\raisebox{.14\height}{\scalebox{.7}{$\textstyle\sum$}}\cr\hidewidth$\textstyle\int$\hidewidth\cr}}
  {\ooalign{\raisebox{.2\height}{\scalebox{.6}{$\scriptstyle\sum$}}\cr$\scriptstyle\int$\cr}}
  {\ooalign{\raisebox{.2\height}{\scalebox{.6}{$\scriptstyle\sum$}}\cr$\scriptstyle\int$\cr}}
}

\def\XXint#1#2#3{{\setbox0=\hbox{$#1{#2#3}{\int}$ }
\vcenter{\hbox{$#2#3$ }}\kern-.6\wd0}}

\def \be  {\begin{equation}}
\def \ee  {\end{equation}}
\def \ba  {\begin{eqnarray}}
\def \ea  {\end{eqnarray}}
\def \baa {\begin{eqnarray*}}
\def \eaa {\end{eqnarray*}}
\def \lab #1 {\label{#1}}

\newcommand\re[1]{(\ref{#1})}
\def\d{\hbox{{d}\kern-.20em\hbox{l}}}

\def \matrix #1 {\left(\begin{array}{cc} #1 \end{array}\right)}

\newcommand \vev [1] {\langle{#1}\rangle}

\newcommand \ket [1] {|{#1}\rangle}

\newcommand{\bit}[1]{\mbox{\boldmath$#1$}}

\newcommand{\ft}[2]{{\textstyle\frac{#1}{#2}}}
\numberwithin{equation}{section}
\begin{document}

\begin{titlepage}

\thispagestyle{empty}

\vspace*{3cm}

\centerline{\large \bf Matrix pentagons}
\vspace*{1cm}

\centerline{\sc A.V.~Belitsky}

\vspace{10mm}

\centerline{\it Department of Physics, Arizona State University}
\centerline{\it Tempe, AZ 85287-1504, USA}

\vspace{2cm}

\centerline{\bf Abstract}

\vspace{5mm}

The Operator Product Expansion for null polygonal Wilson loop in planar maximally supersymmetric Yang-Mills theory runs systematically in terms of multiparticle
pentagon transitions which encode the physics of excitations propagating on the color flux tube ending on the sides of the four-dimensional contour. Their dynamics 
was unravelled in the past several years and culminated in a complete description of pentagons as an exact function of the 't Hooft coupling. In this paper we provide 
a solution for the last building block in this program, the SU(4) matrix structure arising from internal symmetry indices of scalars and fermions. This is achieved by a 
recursive solution of the Mirror and Watson equations obeyed by the so-called singlet pentagons and fixing the form of the twisted component in their tensor decomposition. 
The non-singlet, or charged, pentagons are deduced from these by a limiting procedure.

\end{titlepage}

\setcounter{footnote} 0

\newpage

\pagestyle{plain}
\setcounter{page} 1

%{
%\footnotesize 
%\tableofcontents}

\newpage

\section{Introduction}

A framework for a systematic analysis of the multi-collinear limit of the super Wilson loop in planar $\mathcal{N} = 4$ super Yang-Mills theory on a four-dimensional null polygonal 
contour was proposed in Refs.\ \cite{Alday:2010ku,Basso:2013vsa}. It is akin to the Operator Product Expansion for correlation functions of local operators. The limit of adjacent segments 
of the loop as they approach the same null line introduces curvature field insertions into the Wilson link stretched along this direction. These in turn correspond to excitations on top of the 
Faraday flux tube. Their integrable dynamics was scrutinized in the context of the large-spin limit of high-twist single-trace Wilson operators in the maximally supersymmetric Yang-Mills theory 
\cite{Belitsky:2006en} and is known at any value of the 't Hooft coupling \cite{Basso:2010in}.

\begin{figure}[t]
\begin{center}
\mbox{
\begin{picture}(0,240)(60,0)
\put(0,-140){\insertfig{17}{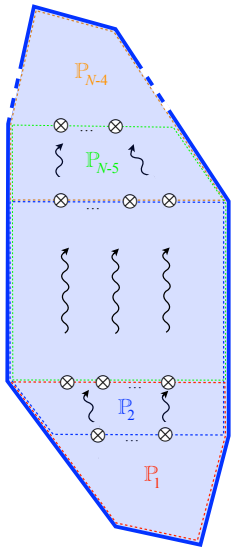}}
\end{picture}
}
\end{center}
\caption{ \label{TesseltationPolygonFig} A tessellation of the polygon into pentagons with a sample set of insertions of flux-tube excitations from the resolution of the identity operator 
on the inner null lines (shown by the $\otimes$ symbols). These propagate from the bottom to the top and interact with each other along the way.}
\end{figure}

A geometric tessellation of the $N$-gon superloop $\mathbb{W}_N$ in null squares introduces the main building block of the formalism, the pentagon $\mathbb{P}$, formed by 
two adjacent squares, yielding the representation
\begin{align}
\mathbb{W}_N
=
\vev{0 | \mathbb{P}_{N-4} \dots \mathbb{P}_2 \mathbb{P}_1 | 0}
\, .
\end{align}
The resolution of the unit operators between sequential pentagons produces the decomposition of the superloop in terms of transition matrix elements of multi-particle flux-tube excitations 
$\ket{{\rm\bf p}_N} \equiv  \ket{ {\rm p}_{1} {\rm p}_{2} \dots {\rm p}_{N} }$ propagating with respective rapidities $\bit{u} = (u_1, u_2, \dots, u_N)$ and interacting on the two-dimensional 
world-sheet of the loop (see Fig.~\ref{TesseltationPolygonFig} for a graphical representation),
\begin{align}
\label{MatrixElementSuperPentagon}
\mathbb{W}_N
=
\SumInt_{N,N', \dots, N''}
\vev{0 | \mathbb{P}_{N-4} | {\rm\bf p}_{N''} (\bit{u}'')}
\dots
\vev{ {\rm\bf p}_{N'} (\bit{u}') | \mathbb{P}_2 | {\rm\bf p}_N (\bit{u})}
\vev{ {\rm\bf p}_N (\bit{u}) | \mathbb{P}_1 | 0 }
\, ,
\end{align}
where we did not display for brevity the $N-5$ accompanying propagation phases or integration measures. The subscripts on the flux-tube excitations cumulatively stand for their Lorentz spins 
and internal symmetry indices. The single-particle spectrum consists of (anti)gluons, scalars, aka holes\footnote{\label{6to6} One can pass to O(6) indices instead making use of the $4 \times 4$ 
off-diagonal blocks $\Sigma_{I, AB}$ of the six-dimensional Dirac matrices in Euclidean metric, such that ${\rm h}^{AB} = \Sigma^{I,AB} {\rm h}^I / \sqrt{2}$. These obey the following involution 
properties $(\Sigma^{I, AB})^\ast = \overline{\Sigma}^I_{AB} \equiv \varepsilon_{ABCD} \Sigma^{I, CD}/2$.}, and (anti)fermions $\ket{{\rm p}} = \ket{\bar{\rm g}}, \ket{{\rm g}}, \ket{{\rm h}^{AB}},  
\ket{\bar\Psi_A}, \ket{\Psi^A},$ which transform in the ${\bf 1}, {\bf 1}, {\bf 6}, {\bf \bar{4}}, {\bf 4}$ of the SU(4) internal symmetry group. In the above formula, the pentagon (or rather the 
superpentagon) $\mathbb{P}$ admits a terminating series in increasing powers of the Grassmann variable $\theta_A$, carrying the index of the antifundamental representation of SU(4),
\begin{align}
\mathbb{P}
=
\mathcal{P} + \theta_A \mathcal{P}^A + \ft1{2!} \theta_A \theta_B \mathcal{P}^{AB} 
+ 
\ft1{3!} \theta_A \theta_B \theta_C \mathcal{P}^{ABC} +  \ft1{4!} \theta_A \theta_B \theta_C \theta_D \mathcal{P}^{ABCD}
\, ,
\end{align}
starting with the singlet $\mathcal{P}$ followed by the SU(4) non-singlet, or charged, operators $\mathcal{P}^A$, $\mathcal{P}^{AB}$ etc.

The matrix elements in \re{MatrixElementSuperPentagon} can be written in the form
\begin{align}
\vev{ {\rm\bf p}_{N'} (\bit{v}) | \mathcal{P}^{A \dots} | {\rm\bf p}_N (\bit{u})}
=
[ \Pi^{A \dots} ]_{N|N'} (\bit{u} | \bit{v}) P (\bit{u}|\bit{v})
\, ,
\end{align}
where the second factor $P (\bit{u}|\bit{v})$ depends on the dynamics of the flux-tube excitations and was the subject of intensive research over the past several years 
\cite{Basso:2013aha,Basso:2014koa,Belitsky:2014sla,Basso:2014nra,Belitsky:2014lta,Belitsky:2015efa,Basso:2014hfa,Basso:2015rta}. In fact, it possesses a factorized form 
in terms of one-to-one particle pentagon transitions \cite{Basso:2013aha,Belitsky:2015efa} as was rigorously demonstrated at leading order in 't Hooft coupling $g$ in the context 
of open (super)spin chains for the flux tube \cite{Belitsky:2014rba,Belitsky:2016fce}.  While the first (matrix) factor $\Pi^{A \dots}_{N|N'}  (\bit{u} | \bit{v})$ encodes information on the 
internal symmetry indices and enjoys rational dependence on differences of particles' rapidities. It is independent of $g$ and is thus purely kinematical in origin. It is the focus of the 
present work. At this point, it is worth pointing out that both of the above facts are, in principle, conjectures. However, they withstood all tests conducted to date against explicit data 
on scattering amplitudes made available by other means and methods. For the case at hand, the uniqueness of matrix part was again verified purely empirically as will be further
discussed later.

%%%%%%%%%%%%%%%%%%%%%%%%%%%%%%%%%%%%%%%%%%%%%%%%%%%%%%%%%%%%%%%%%%%%%
%            Figure
%%%%%%%%%%%%%%%%%%%%%%%%%%%%%%%%%%%%%%%%%%%%%%%%%%%%%%%%%%%%%%%%%%%%%
%%%%%%%%%%%%%%%%%%%%%%%%%%%%%%%%%%%%%%%%%%%%%%%%%%%%%%%%%%%%%%%%%%%%%

Our subsequent presentation is organized as follows. In the next section, we start with the matrix elements of the singlet pentagon operator involving only holes and provide a systematics 
procedure for construction of all terms in its tensor decomposition which is based on the solution of Mirror and Watson equations obeyed by matrix pentagons. The seed for this recursion 
is provided by just one component which requires absolute fixing. Next, we move on to the purely fermion helicity-preserving transitions. Then we conclude with mixed fermion-hole singlet 
pentagonsand finally with the transition involving all charged excitations. In Sect.\ \ref{MovingSection}, we address the question of moving excitations from the initial to final state, giving an 
effective set of rules for fermions which lack a simple one-particle mirror transformation. We then demonstrate how to deduce the non-singlet transitions in Sect.\ \ref{NonsingletSection} from 
the ones we just computed. We construct integrands of polygon loops in the flux-tube representation and verify our findings by comparing them with the integral representation suggested in 
Ref.\ \cite{Basso:2014jfa,Basso:2015uxa} for the hexagonal Wilson loop in Sect.\ \ref{IntegrandTestSection}, finding agreement. In Appendix \ref{ExamplesAppendix} we give a few examples 
of tensors with small number of particles, leaving the rest to the accompanying Mathematica notebook that contains routines for automatic solution of systems of Mirror and Watson equations, 
testing results against integral representation of the hexagon and limiting procedure to obtain all transition matrices from the minimal set considered in this paper.

\section{Singlet pentagons}
\label{SingletSection}

To begin with, we address the matrix structure of the lowest Grassmann component $\mathcal{P}$ in the expansion of the superpentagon $\mathbb{P}$. We will discuss in
turn three cases of increasing complexity from purely hole transition matrix elements passing to purely fermionic ones and finally addressing their mixed states.

\subsection{Hole matrices}

We start with a comment. The singlet pentagon operator itself obviously does not carry any SU(4) indices, so its matrix elements can have a total even number of holes shared between the 
initial and final states. In this section we provide a solution to the diagonal $N$-to-$N$ case. The particle number-changing transitions can be deduced from this one making use of the 
known double Wick, aka mirror, transformation properties which allow one to move excitations between different sides of the pentagon.

With this in mind, let us introduce transitions from the initial state of $N$ scalars carrying rapidities $\bit{u} = (u_1, \dots, u_N)$ and O(6) indices\footnote{We will find useful using the 
SU(4) indices instead when discussing mixed matrix elements. For the time being the O(6) conventions are the most economical.} $\bit{I} = (I_1, \dots, I_N)$, cumulatively called 
${\rm\bf h}^{\bit{\scriptstyle I}} (\bit{u})$, to the final state of $N$ scalars ${\rm\bf h}^{\bit{\scriptstyle J}} (\bit{v})$,
\begin{align}
\label{MultiHoleTensorPentagon}
P^{\bit{\scriptstyle I} | \bit{\scriptstyle J}} (\bit{u} | \bit{v}) 
= 
\vev{ {\rm\bf h}^{\bit{\scriptstyle J}} (\bit{v}) | \mathcal{P} | {\rm\bf h}^{\bit{\scriptstyle I}} (\bit{u})}
\, .
\end{align}
The above pentagons can be cast in the form of a scalar factor accompanied by an O(6) tensor
\begin{align}
\label{HoleHoleTensorPentagon}
P^{\bit{\scriptstyle I} | \bit{\scriptstyle J}} (\bit{u} | \bit{v})
=
\Pi^{\bit{\scriptstyle I} | \bit{\scriptstyle J}} (\bit{u} | \bit{v})
P_{{\rm\bf h} | {\rm\bf h}} (\bit{u} | \bit{v}) 
\, .
\end{align}
Here $P_{{\rm\bf h} | {\rm\bf h}} (\bit{u} | \bit{v})$ contains dynamical information about the transition of $N$-to-$N$ hole states through the dependence on the 't 
Hooft coupling. It was shown to admit a factorized form in terms of two-particle pentagons \cite{Basso:2013aha,Belitsky:2015efa,Belitsky:2014rba,Belitsky:2016fce}
\begin{align}
\label{DynamicalHoleNpentagon}
P_{{\rm\bf h} | {\rm\bf h}} (\bit{u} | \bit{v}) 
=
\frac{\prod\limits_{i,j}^N P_{\rm h|h} (u_i|v_j)}{\prod\limits_{i>j}^N P_{\rm h|h} (u_i | u_j) \prod\limits_{i<j}^N P_{\rm h|h} (v_i | v_j)}
\, .
\end{align}

%%%%%%%%%%%%%%%%%%%%%%%%%%%%%%%%%%%%%%%%%%%%%%%%%%%%%%%%%%%%%%%%%%%%%
%            Figure
%%%%%%%%%%%%%%%%%%%%%%%%%%%%%%%%%%%%%%%%%%%%%%%%%%%%%%%%%%%%%%%%%%%%%
\begin{figure}[t]
\begin{center}
\mbox{
\begin{picture}(0,120)(190,0)
\put(20,-250){\insertfig{17}{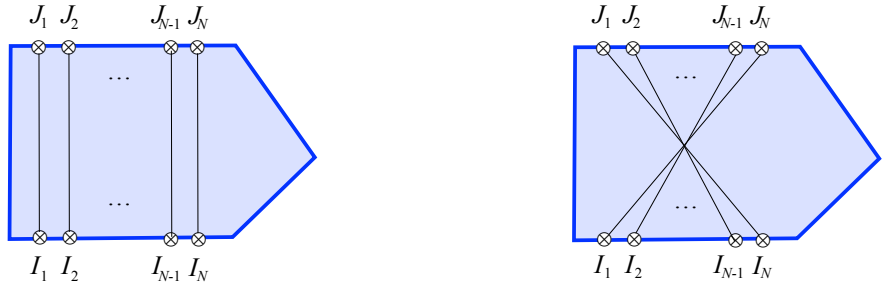}}
\end{picture}
}
\end{center}
\caption{ \label{PefectPairingFig} Two contributions out of $(2N-1)!!$ of perfect pairings of O(6) indices of holes which are displayed in Eq.\ \re{HoleHoleTensor}. The 
picture on the right shows the twisted graph which corresponds to the seed rational function for the recursive solution of the defining equations.}
\end{figure}
%%%%%%%%%%%%%%%%%%%%%%%%%%%%%%%%%%%%%%%%%%%%%%%%%%%%%%%%%%%%%%%%%%%%%

As was already stated in the Introduction, the matrix $\Pi^{\bit{\scriptstyle I} | \bit{\scriptstyle J}} (\bit{u} | \bit{v})$ does not depend on the coupling constant and,
as a function of the rapidity variables, it enjoys dependence only through their differences and is purely rational in nature. Its tensor decomposition runs over $(2N-1)!!$ 
perfect pairings of all indices\footnote{We adopt the numbering scheme that naturally emerges from a pairing routine in the accompanying Mathematica notebook.}
\begin{align}
\label{HoleHoleTensor}
\Pi^{\bit{\scriptstyle I} | \bit{\scriptstyle J}} (\bit{u} | \bit{v}) 
=
\dots
&
+ 
\delta^{I_1 J_1} \delta^{I_2 J_2} \delta^{I_3 J_3} \dots \delta^{I_N J_N} \pi_{[(2N-1)!! + 1]/2} (\bit{u} | \bit{v}) 
+
\dots\\
&
+ 
\delta^{I_1 J_N} \delta^{I_2 J_{N-1}} \delta^{I_3 J_{N-2}} \dots \delta^{I_N J_1} \pi_{(2N-1)!!} (\bit{u} | \bit{v})
\, ,
\nonumber
\end{align}
shown schematically in Fig.~\ref{PefectPairingFig}. The last matrix structure corresponds to the twisted graph, i.e., when the ordering on all of the sites on the top is completely 
reversed, i.e., $(1 2 3 \dots N) \to (N \dots 3 2 1)$. It will play a distinguished role in our consideration. In principle, one could introduce extra tensor structures involving odd number
of SO(6) Levi-Civita symbols for each sextet of holes. However, solution to Mirror and Watson equation combined with Bose symmetry do not yield nontrivial solutions for the
corresponding structures. Thus, they will be ignored in what follows.

%%%%%%%%%%%%%%%%%%%%%%%%%%%%%%%%%%%%%%%%%%%%%%%%%%%%%%%%%%%%%%%%%%%%%
%            Figure
%%%%%%%%%%%%%%%%%%%%%%%%%%%%%%%%%%%%%%%%%%%%%%%%%%%%%%%%%%%%%%%%%%%%%
\begin{figure}[p]
\begin{center}
\mbox{
\begin{picture}(0,500)(180,0)
\put(11,190){\insertfig{17}{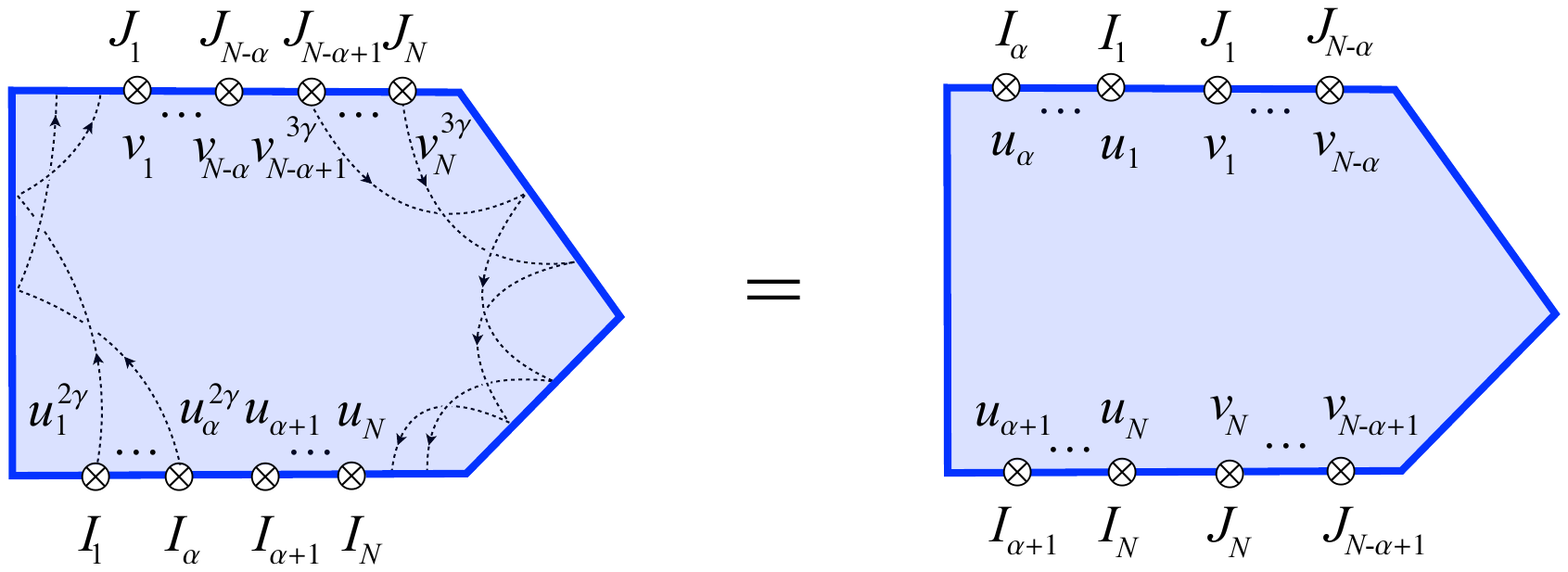}}
\put(10,70){\insertfig{17}{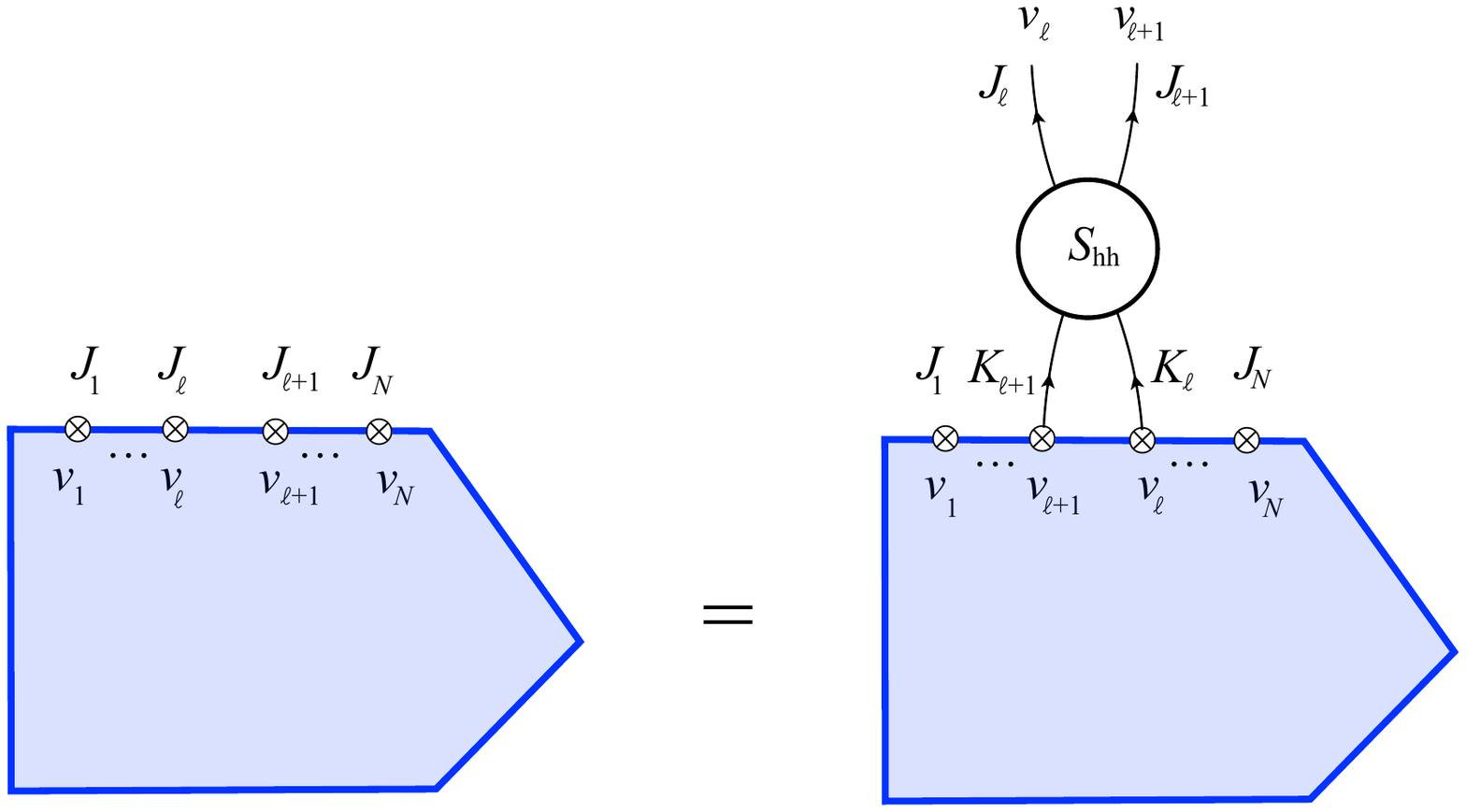}}
\put(0,-190){\insertfig{17}{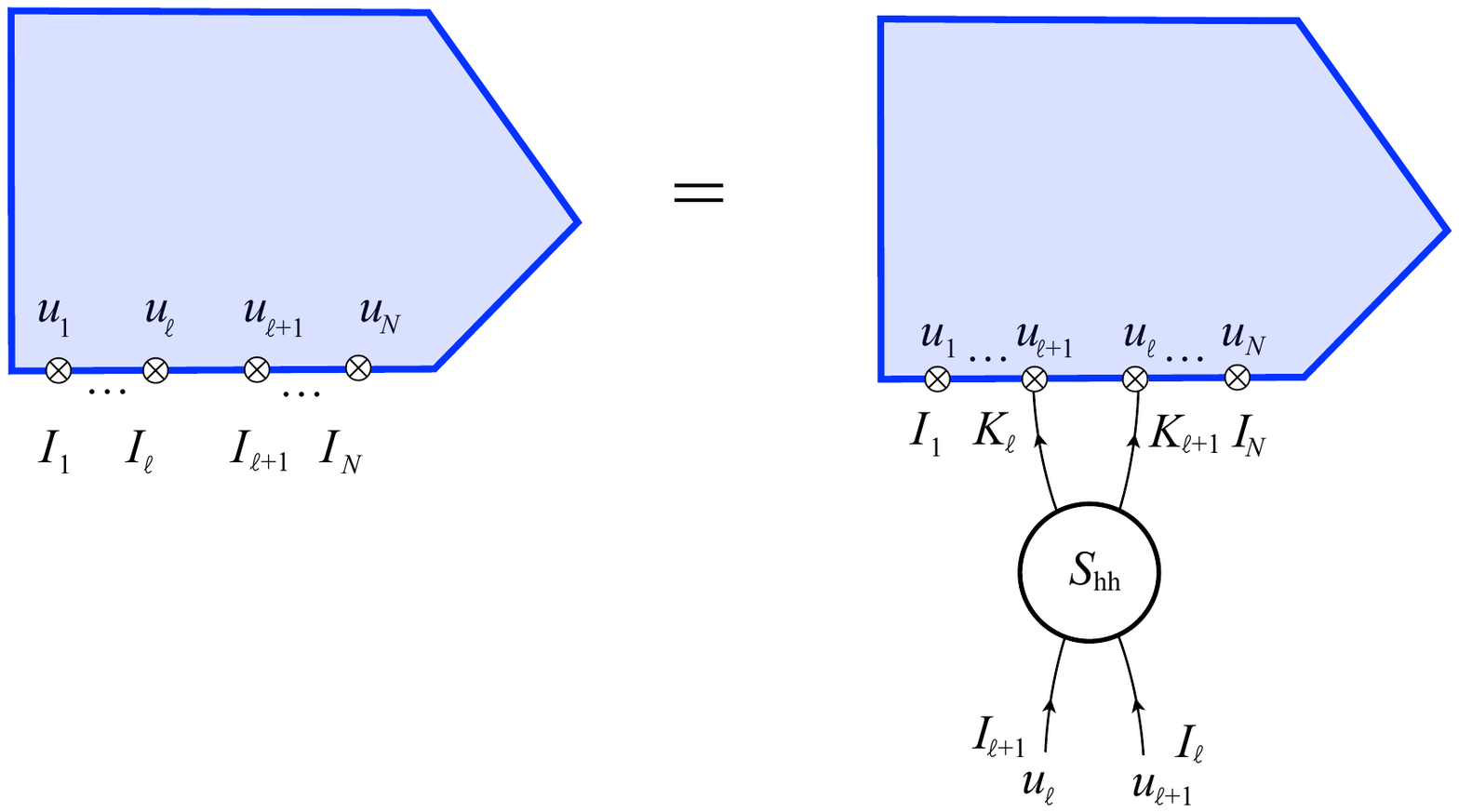}}
\end{picture}
}
\end{center}
\caption{ \label{MirrorAndWatson} From top to bottom: Graphical representation of the Mirror and Watson equations for (top and bottom) hole flux-tube 
excitations.}
\end{figure}
%%%%%%%%%%%%%%%%%%%%%%%%%%%%%%%%%%%%%%%%%%%%%%%%%%%%%%%%%%%%%%%%%%%%%

\subsubsection{Solution to Mirror and Watson equations}
\label{MirrorWatsonEqSolutionSection}

The matrix pentagon \re{HoleHoleTensorPentagon} obeys a system of defining relations. It is formed by the Mirror and Watson equations. The first of this kind emerges from the 
invariance of the flux-tube background with respect to the double Wick rotation \cite{Alday:2010ku}, which allows one to interchange space and time variables on the two-dimensional 
worldsheet of the loop. From the point of view of the hole excitation, this interchanges the energy and momentum in its dispersion relation. As a function of the rapidity variable, an 
analytic continuation that accomplishes this goal was found in Ref.\ \cite{Basso:2011rc}. For the hole-to-hole pentagon transition, it takes the following form \cite{Basso:2013aha}
\begin{align}
P_{\rm h|h} (u|v^\gamma) = P_{\rm h|h} (v|u)
\, ,
\end{align}
where $\gamma$ stands for the aforementioned path in the complex rapidity plane. In fact, since $P_{\rm h|h} (u|v)$ is a meromorphic function of rapidities with an infinite number of
cuts equidistantly spaced along the imaginary axis $[-2g + i (k +\ft12), 2g + i (k + \ft12)]$, with $k \in \mathbb{Z}$, the continuation $v^\gamma = v + i$ implies going through
the lowest cut in the upper half-plane and passing to another (mirror) Riemann sheet \cite{Basso:2011rc,Basso:2013pxa}. Multiple application of the mirror transformation to the same 
excitation allows one to move it from the initial to the final state, yielding a creation form factor \cite{Basso:2013aha}
\begin{align}
\label{2GammaHole}
P_{\rm h|h} (u^{2 \gamma}|v) = \frac{R_{\rm hh} (u, v)}{P_{\rm h|h} (u|v)}
\, , \qquad
R_{\rm hh} (u, v)
=
\frac{1}{(u|v)_1 (u|v)_2}
\, .
\end{align} 
It is related to the inverse of the original transition pentagon up to an overall rational function $R_{\rm hh} (u, v)$ of hole rapidities. Here and below, we use the notation
\begin{align}
(u|v)_\sigma \equiv u - v + i \sigma
\end{align}
to make expressions more compact. Obviously, $(u|v)_\sigma = - (v|u)_{- \sigma}$. 

For the multi-hole matrix pentagon \re{MultiHoleTensorPentagon}, moving $\alpha$ excitations from the top to bottom and the same number of the bottom ones to the top, say, in the 
clockwise direction, yields the same object but with accordingly changed rapidities and O(6) matrix structure. The Mirror equations, shown diagrammatically by the top panel 
in Fig.\ \ref{MirrorAndWatson}, then read
\begin{align}
&
P^{\bit{\scriptstyle I} | \bit{\scriptstyle J}} (\bit{u} + 2 i \bit{\alpha} | \bit{v} + 3 i \bit{\bar\alpha})
\\
&
=
P^{{\scriptstyle I_{\alpha + 1}, \dots, I_N, J_N, \dots, J_{N - \alpha + 1}} | {\scriptstyle I_\alpha, \dots, I_1, J_1, \dots, J_{N - \alpha}}}
(u_{\alpha + 1}, \dots, u_N, v_N, \dots, v_{N - \alpha + 1} | u_\alpha, \dots, u_1, v_1, \dots, v_{N - \alpha})
\, . \nonumber
\end{align}
Here we introduced vectors $\bit{\alpha} $ and $\bit{\bar\alpha} $ with unit components
\begin{align}
\bit{\alpha} = (\overbrace{1, \dots, 1}^\alpha, 0, \dots 0)
\, , \qquad
\bit{\bar\alpha} = (0, \dots, 0, \overbrace{1, \dots, 1}^\alpha) 
\end{align}
and their length $\alpha = |\bit{\alpha}| = |\bit{\bar\alpha}|$.

The Watson equations can be written either for the initial or final state. They are, respectively,
\begin{align}
\label{WatsonEqBottom}
P^{\bit{\scriptstyle I} | \bit{\scriptstyle J}} (\bit{u} | \bit{v})
&
=
S^{{\scriptstyle I_{\ell+1} I_\ell} | {\scriptstyle K_{\ell + 1} K_\ell}} (u_\ell, u_{\ell + 1})
P^{{\scriptstyle I_1, \dots, K_{\ell + 1}, K_\ell, \dots,  I_N} | \bit{\scriptstyle J}} (u_1, \dots, u_{\ell + 1}, u_\ell, \dots, u_N | \bit{v})
\, , \\
\label{WatsonEqTop}
P^{\bit{\scriptstyle I} | \bit{\scriptstyle J}} (\bit{u} | \bit{v})
&
=
S^{{\scriptstyle K_{\ell + 1} K_\ell} | {\scriptstyle J_{\ell+1} J_\ell}} (v_{\ell + 1}, v_\ell)
P^{\bit{\scriptstyle I} | {\scriptstyle J_1, \dots, K_{\ell + 1}, K_\ell, \dots,  J_N}} (\bit{u} | v_1, \dots, v_{\ell + 1}, v_\ell, \dots, v_N)
\, ,
\end{align}
with $1 \leq \ell \leq N-1$ and where the S-matrix for scattering of the sextet of scalar excitations
\begin{align}
S^{I_1 I_2 | J_1 J_2} (u, v)
=
S_{\rm hh} (u,v)
\left[
\delta^{I_1 J_1} \delta^{I_2 J_2} s^{(1)}_{\rm hh} (u, v)
+
\delta^{I_1 J_2} \delta^{I_2 J_1} s^{(2)}_{\rm hh} (u, v)
+
\delta^{I_1 I_2} \delta^{J_1 J_2} s^{(3)}_{\rm hh} (u, v)
\right]
\, ,
\end{align}
differs from the Zamolodchikovs' O(6) matrix by the overall dynamical phase $S_{\rm hh} (u,v)$ that encodes information on the flux-tube background as a function of the
't Hooft coupling. The nested Bethe Ansatz uniquely determines the rational factors in front of the identity, permutation and annihilation tensors \cite{Zamolodchikov:1977nu}
\begin{align}
s^{(1)}_{\rm hh} (u, v)
=
\frac{u - v}{u - v - i}
\, , \qquad
s^{(2)}_{\rm hh} (u, v)
=
\frac{-i}{u - v - i}
\, , \qquad
s^{(3)}_{\rm hh} (u, v)
=
\frac{i (u - v)}{(u - v - i)(u - v - 2i)}
\, ,
\end{align}
respectively. The two equations, \re{WatsonEqBottom} and \re{WatsonEqTop}, contain identical information, so only one of them provides an independent set of 
relations between  $\pi$-functions. As a consequence, one is free to choose one of the above for the recursive solution of form factors in question.

It is important to realize that the Watson equation alone is not sufficient in general to determine all coefficients recursively. One has to rely on the Mirror equation as
well to express all $\pi$'s in terms of just one, $\pi_{(2N-1)!!} (\bit{u} | \bit{v})$, in front of the twisted matrix structure. This last one has to be absolutely fixed and
the most stringent constraint for it arises from the Mirror equations. The latter are specific to the flux-tube dynamics of the maximally supersymmetric Yang-Mills 
theory and, therefore, cannot be used in a generic form factor program. They read
\begin{align}
&\frac{
\pi_{(2N-1)!!} (u_{\alpha + 1}, \dots, u_N, v_N, \dots, v_{N - \alpha + 1} | u_\alpha, \dots, u_1, v_1, \dots, v_{N - \alpha})
}{
\pi_{(2N-1)!!} (\bit{u} + 2 i \bit{\alpha} | \bit{v} + 3 i \bit{\bar\alpha} )
}
\\
&=
\frac{
\prod\limits_{j_1 = 1}^\alpha \prod\limits_{k_1 = \alpha + 1}^N (u_{j_1} | u_{k_1})_0 (u_{j_1} | u_{k_1})_1
\prod\limits_{j_2 = N - \alpha + 1}^N \prod\limits_{k_2 = 1}^{N - \alpha} (v_{k_2} | v_{j_2})_{-1} (v_{k_2} | v_{j_2})_{-2}
}{
\prod\limits_{j_1 = 1}^\alpha \prod\limits_{k_1 = 1}^{N - \alpha} (u_{j_1} | v_{k_1})_1 (u_{j_1} | v_{k_1})_2
\prod\limits_{j_2 = N - \alpha + 1}^N \prod\limits_{k_2 = \alpha + 1}^{N} (u_{k_2} | v_{j_2})_{-1} (u_{k_2} | v_{j_2})_{-2}
}
\, . \nonumber
\end{align} 
The origin of the rational function in the right-hand side is traced back to the product of $R_{\rm h|h}$ coefficients in Eq.\ \re{2GammaHole}. The solution to these equations 
is given by the quotient of same degree polynomials in rapidity variables in the numerator and denominator,
\begin{align}
\pi_{(2N-1)!!} (\bit{u} | \bit{v})
=
\frac{
\prod\limits_{k_1=1}^{N-1} \prod\limits_{j_1=1}^{N-k_1} (u_{j_1} | v_{k_1})_0
\prod\limits_{j_2=2}^{N} \prod\limits_{k_2=j_2}^{N} (u_{k_2} | v_{N - j_2 + 2})_1
}{
\prod\limits_{j_1=1}^{N-1} \prod\limits_{k_1 = j_1 + 1}^{N-1} (u_{j _1} | u_{k_1})_{-1} (v_{j_1} | v_{k_1})_1
}
\, .
\end{align}
The correctness of this solution was verified by means of dedicated perturbative analyses for low number of particles, see, e.g., Refs. \cite{Basso:2014koa,Basso:2014hfa}. 
We provide an explicit example for $2 \to 2$ and $3 \to 3$ transitions in Appendix \ref{ExamplesAppendixScalars}. Expressions for larger number of particles are prohibitively 
long to be displayed explicitly in the paper and are more suitable in a symbolic form of the accompanying Mathematica notebook.

\subsection{Fermion matrices}

Let us continue with pentagon transitions involving only fermions, namely, the ones corresponding to $N$ fermions in the initial state and the same number 
of antifermions in the final state
\begin{align}
\label{FermionicPentagons}
P^{\bit{\scriptstyle A} |}{}_{\bit{\scriptstyle B}} (\bit{u}|\bit{v})
= 
\vev{ \bf{\bar\Psi}_{\bit{\scriptstyle B}} (\bit{v}) | \mathcal{P} | \bf{\Psi}^{\bit{\scriptstyle A}} (\bit{u}) }
=
\Pi^{\bit{\scriptstyle A}|}{}_{\bit{\scriptstyle B}} (\bit{u} | \bit{v})
P_{{\bf \Psi} | {\bf \Psi}} (\bit{u} | \bit{v})
\, .
\end{align}
These correspond to the helicity-preserving matrix elements. Notice that the SU(4) symmetry also allows for transitions involving quartets of (anti)fermions in addition to
the excitations already present in the in- and out-states due to possibility to carry internal symmetry group indices by the four-dimensional Levi-Civita tensor, however, these 
will be obtained from the ones we are about to analyze by taking a particular limit. 

The decomposition in independent tensors is straightforward and arises from the pairwise contraction of the bottom and top indices with Kronecker symbols and $N!$ permutations 
of either the top or bottom positions,
\begin{align}
\label{FermionFermionTensor}
\Pi^{\bit{\scriptstyle A} |}{}_{\bit{\scriptstyle B}} (\bit{u}|\bit{v})
=
\delta^{A_1}_{B_1} \delta^{A_2}_{B_2} \dots \delta^{A_N}_{B_N} \pi_1 (\bit{u} | \bit{v})
+ \dots
+
\delta^{A_1}_{B_N} \delta^{A_2}_{B_{N-1}} \dots \delta^{A_N}_{B_1} \pi_{N!} (\bit{u} | \bit{v})
\, ,
\end{align}
with displayed terms shown graphically in Fig.\ \ref{PefectPairingFig}.

Fermions do not enjoy a simple mirror transformation \cite{Basso:2014koa} so we do not have an equation to fix the twisted component. However, by analogy with 
the case of scalars discussed in the previous section, we anticipate that the rational function should differ from it only marginally, i.e., possibly by the imaginary shifts 
due to different helicity of the excitations involved if at all. In fact, we conjecture the $\pi_{N!}$ to take the form 
\begin{align}
\label{TwistedFermionPi}
\pi_{N!} (\bit{u} | \bit{v})
=
\frac{
\prod\limits_{k_1=1}^{N-1} \prod\limits_{j_1=1}^{N-k_1} (u_{j_1} | v_{k_1})_0
\prod\limits_{j_2=2}^{N} \prod\limits_{k_2=j_2}^{N} (u_{k_2} | v_{N - j_2 + 2})_1
}{
\prod\limits_{j_1=1}^{N-1} \prod\limits_{k_1 = j_1 + 1}^{N-1} (u_{j _1} | u_{k_1})_{-1} (v_{j_1} | v_{k_1})_1
}
\, . 
\end{align}
We want to emphasize that this form is intrinsic to the flux-tube dynamics.

This seed provides the solution for the matrix structure in question since all functions accompanying other structures can be extracted making use of the Watson 
equations alone, contrary to the scalar sector where the number of independent components is much higher and one has to rely on additional relations emerging from
the Mirror equations. The Watson equations for the fermion take the same form as Eqs.\ \re{WatsonEqBottom} -- \re{WatsonEqTop} with obvious replacements of 
O(6) indices on the bottom/top with covariant/contravariant SU(4) indices and the fermion-fermion scattering matrix being
\begin{align}
S^{A_1 A_2}_{B_1 B_2} (u, v) 
= 
S_{\Psi \Psi} (u,v)
\left[
\delta^{A_1}_{B_1} \delta^{A_2}_{B_2} s_{\Psi\Psi}^{(1)} (u,v)
+
\delta^{A_1}_{B_2} \delta^{A_2}_{B_1} s_{\Psi\Psi}^{(2)} (u,v)
\right]
\, ,
\end{align}
where the component of the $R$-matrix are \cite{Berg:1977dp}
\begin{align}
s^{(1)}_{\Psi\Psi} (u, v)
=
\frac{u - v}{u - v - i}
\, , \qquad
s^{(2)}_{\Psi\Psi} (u, v)
=
\frac{-i}{u - v - i}
\, .
\end{align}
Due to a much smaller number of independent structures in Eq.\ \re{FermionFermionTensor}, recursive solution to Watson equations allow one to find all $\pi$'s starting 
with \re{TwistedFermionPi}. We give an example in Appendix \ref{ExamplesAppendixFermions}. All other multiparticle pentagons can be found analogously making use 
of the automatic solver in the accompanying notebook.

\subsection{Mixed matrices}

Last but not least, we address the case when both holes and (anti)fermions are present in the transition. We start with holes and antifermions, first, and then add fermions to
the mix. 

Namely, the $N$ holes to $2N$ antifermion transitions,
\begin{align}
P^{\bit{\scriptstyle AB} |}{}_{\bit{\scriptstyle C}} (\bit{u}|\bit{v})
= 
\vev{ \bit{\bar\Psi}_{\bit{\scriptstyle C}} (\bit{v}) | \mathcal{P} | {\rm\bf h}^{\bit{\scriptstyle AB}} (\bit{u}) }
\, ,
\end{align}
where the rapidity arrays are $N$, $\bit{u} = (u_1, \dots, u_N)$, and $2N$, $\bit{v} = (v_1, \dots, v_{2N})$, dimensional, respectively, and the sets of the SU(4) indices in the 
defining representation having the same lengths, $\bit{A} = (A_1, \dots, A_N)$, $\bit{B} = (B_1, \dots, B_N)$ and $\bit{C} = (C_1, \dots, C_{2N})$. The above matrix element 
factorizes as before
\begin{align}
P^{\bit{\scriptstyle AB} |}{}_{\bit{\scriptstyle C}} (\bit{u}|\bit{v})
=
\Pi^{\bit{\scriptstyle AB}|}{}_{\bit{\scriptstyle C}} (\bit{u} | \bit{v})
P_{{\rm\bf h} | {\bf \Psi}} (\bit{u} | \bit{v}) 
\, .
\end{align}
Here $P_{{\rm\bf h} | {\bf \Psi}} (\bit{u} | \bit{v})$ admits again the form
\begin{align}
P_{{\rm\bf h} | {\bf \Psi}} (\bit{u} | \bit{v}) 
=
\frac{\prod\limits_{i, j}^{N, 2N} P_{\rm h|\Psi} (u_i|v_j)}{\prod\limits_{i>j}^N P_{\rm h|h} (u_i | u_j) \prod\limits_{i<j}^{2N} P_{\Psi|\Psi} (v_i | v_j)}
\, .
\end{align}
The decomposition of $\Pi^{\bit{\scriptstyle AB}|}{}_{\bit{\scriptstyle C}} (\bit{u} | \bit{v})$ into independent tensors is accomplished in the same manner as for the purely fermionic 
transitions discussed above, i.e., generating $2N!$ different pairings. However, this time one has to impose additional constraints for antisymmetry of $N$ pairs of $\bit{A}$ and 
$\bit{B}$ indices. This yields a total number of $2N!/2^N$ independent structures,
\begin{align}
\label{HoleFermionTensor}
\Pi^{\bit{\scriptstyle AB}|}{}_{\bit{\scriptstyle C}} (\bit{u} | \bit{v})
&
=
\delta^{[A_1}_{C_1} \delta^{B_1]}_{C_2} 
\dots
\delta^{[A_N}_{C_{2N-1}} \delta^{B_N]}_{C_{2N}} 
\pi_1  (\bit{u} | \bit{v})
+
\dots
\nonumber\\
&
+
\delta^{[A_N}_{C_1} \delta^{B_N]}_{C_2} 
\dots
\delta^{[A_1}_{C_{2N-1}} \delta^{B_1]}_{C_{2N}}
\pi_{2N!/2^N}  (\bit{u} | \bit{v})
\, .
\end{align}
As in the purely fermionic case, there are no closed mirror equations for the amplitude in question. So we will conjecture the twisted component again.
It will take the form of the previous two cases, with a generalization to account for twice the number of rapidities on the top of the pentagon. Basically, we 
lump them up in nearest-neighbor pairs starting with the first position and double the number of rational factors in the numerator. Taking into account different 
values of helicity which result in half-integer imaginary shifts, we find
\begin{align}
\pi_{2N!/2^N} (\bit{u} | \bit{v})
=
\frac{
\prod\limits_{k_1=1}^{N-1} \prod\limits_{j_1=1}^{N-k_1} (u_{j_1} | v_{2 k_1 - 1})_{-1/2} (u_{j_1} | v_{2 k_1})_{-1/2}
\prod\limits_{j_2=2}^{N} \prod\limits_{k_2=j_2}^{N} (u_{k_2} | v_{2 N - 2 j_2 + 3})_{3/2} (u_{k_2} | v_{2 N - 2 j_2 + 4})_{3/2}
}{
\prod\limits_{j_1=1}^{N-1} \prod\limits_{k_1 = j_1 + 1}^{N-1} (u_{j _1} |u_{k_1})_{-1}  (u_{j _1} | u_{k_1})_{-2} 
\prod\limits_{j_2=1}^{2N-1} \prod\limits_{k_2 = j_2 + 1}^{2N-1} (v_{j_2} | v_{k_2})_1
}
\, . 
\end{align}
The remaining functions in the tensor decomposition \re{HoleFermionTensor} arise from this by repeated use of $(2N-1)$ final-state Watson equations involving
fermionic S-matrices of the previous subsection. This is demonstrated on a simple example in Appendix \ref{ExamplesAppendixHolesFermions}, with higher particle 
number cases deferred to the accompanying file.

%%%%%%%%%%%%%%%%%%%%%%%%%%%%%%%%%%%%%%%%%%%%%%%%%%%%%%%%%%%%%%%%%%%%%
%            Figure
%%%%%%%%%%%%%%%%%%%%%%%%%%%%%%%%%%%%%%%%%%%%%%%%%%%%%%%%%%%%%%%%%%%%%
\begin{figure}[t]
\begin{center}
\mbox{
\begin{picture}(0,160)(100,0)
\put(0,-270){\insertfig{20}{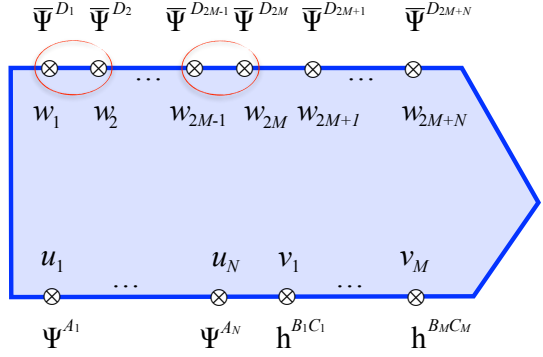}}
\end{picture}
}
\end{center}
\caption{ \label{MixedMatrixPicture} Distribution of excitations on the mixed pentagon. The rapidities of the first $2M$ antifermions on the top are 
lumped in pairs such that the twisted structure follows the same pattern as purely hole/fermion on up to different shift assignments.}
\end{figure}
%%%%%%%%%%%%%%%%%%%%%%%%%%%%%%%%%%%%%%%%%%%%%%%%%%%%%%%%%%%%%%%%%%%%%

Finally, it is left to consider all types of excitations with isotopic indices residing on the contour. The simplest case, that is the basis for all other possibilities, is of $N$ 
fermions with rapidities $\bit{u} = (u_1, \dots, u_{N})$ and $M$ holes with rapidities $\bit{v} = (v_1, \dots, v_{M})$ on the bottom and $N + 2M$ antifermions with
rapidities $\bit{w} = (w_1, \dots, w_{N + 2M})$ on the top,
\begin{align}
P^{\bit{\scriptstyle ABC} |}{}_{\bit{\scriptstyle D}} (\bit{u}, \bit{v} | \bit{w})
= 
\vev{ \bit{\bar\Psi}_{\bit{\scriptstyle D}} (\bit{w}) | \mathcal{P} | \bit{\Psi}^{\bit{\scriptstyle A}} (\bit{u}) {\rm\bf h}^{\bit{\scriptstyle BC}} (\bit{v}) }
\, .
\end{align}
Again the SU(4) matrix in the factorized expression
\begin{align}
P^{\bit{\scriptstyle ABC} |}{}_{\bit{\scriptstyle D}} (\bit{u}, \bit{v} | \bit{w})
=
\Pi^{\bit{\scriptstyle ABC} |}{}_{\bit{\scriptstyle D}} (\bit{u}, \bit{v} | \bit{w})
P_{\rm\bf \Psi h | \Psi} (\bit{u}, \bit{v} | \bit{w})
\end{align}
admits the form
\begin{align}
\label{HoleFermionAntiFermionTensor}
&
\Pi^{\bit{\scriptstyle ABC} |}{}_{\bit{\scriptstyle D}} (\bit{u}, \bit{v} | \bit{w})
\\
&
=
\delta^{A_1}_{D_1} \dots \delta^{A_{N}}_{D_{N}} 
\delta^{[ B_1}_{D_{N + 1}} \delta^{C_1]}_{D_{N + 2}}
\dots
\delta^{[ B_{M}}_{D_{N + 2 M - 1}} \delta^{C_{M}]}_{D_{N + 2 M}} \pi_1  (\bit{u}, \bit{v} | \bit{w})
+
\dots
\nonumber\\
&
+
\delta^{A_1}_{D_{N + 2 M}} \dots \delta^{A_{N}}_{D_{2 M + 1}} 
\delta^{[ B_1}_{D_{2 M}} \delta^{C_1]}_{D_{2 M - 1}}
\dots
\delta^{[ B_{M}}_{D_2} \delta^{C_{M}]}_{D_1}
\pi_{(N+2M)!/2^{M}}  (\bit{u}, \bit{v} | \bit{w})
\, . \nonumber
\end{align}
The dynamical term has the same structure as earlier in terms of one-to-one pentagons
\begin{align}
P_{\rm\bf \Psi h | \Psi} (\bit{u}, \bit{v} | \bit{w})
=
\frac{
\prod\limits_{i,j}^{N, N + 2M} P_{\Psi | \Psi} (u_i|w_j) 
\prod\limits_{i,j}^{M, N + 2M} P_{{\rm h}|\Psi} (v_i|w_j) 
}{
\prod\limits_{i<j}^{N + 2M} P_{\Psi | \bar\Psi} (w_i | w_j)
\prod\limits_{i>j}^{N} P_{\Psi | \bar\Psi} (u_i | u_j)
\prod\limits_{i>j}^{M} P_{{\rm h} | {\rm h}} (v_i | v_j)
\prod\limits_{i,j}^{M,N} P_{{\rm h} | \Psi} (v_i | u_j)
}
\, .
\end{align}
The twisted function in the matrix part now reads
\begin{align}
\pi_{(N+2M)!/2^{M}} (\bit{u}, \bit{v} | \bit{w})
=
\frac{
\mathcal{N}_1 (\bit{u}, \bit{v} | \bit{w}) \mathcal{N}_2 (\bit{u}, \bit{v} | \bit{w})
}{
\mathcal{D} (\bit{u}, \bit{v} | \bit{w})
}
\, ,
\end{align}
where
\begin{align}
\mathcal{N}_1 (\bit{u}, \bit{v} | \bit{w}) 
&
=
\prod\limits_{k_1 = 2M + 1}^{N + 2M - 1} 
\prod\limits_{j_1 = 1}^{N + 2M - k_1} 
(u_{j_1} | w_{k_1})_{0}
\nonumber\\
&\times
\prod\limits_{k_2 = 1}^{M} 
\prod\limits_{j_2 = 1}^{M - k_2} 
(v_{j_2} | w_{2 k_2 - 1})_{-1/2} (v_{j_2} | w_{2 k_2})_{-1/2} 
\prod\limits_{j_3 = 1}^{N} 
(u_{j_3} | w_{2 k_2 - 1})_{0} (u_{j_3} | w_{2 k_2})_{0} 
\, , \\
\mathcal{N}_2 (\bit{u}, \bit{v} | \bit{w}) 
&
=
\prod\limits_{j_1 = 0}^{N-1} 
\prod\limits_{k_1 = j_1}^{N-2} 
(u_{k_1 + 2} | w_{N + 2M - j_1})_{1}
\prod\limits_{k_2 = 1}^{M} 
(v_{k_2} | w_{N + 2M - j_1})_{3/2} 
\nonumber\\
&\times
\prod\limits_{j_2 = 0}^{M - 2} 
\prod\limits_{k_3 = j_2}^{M - 2} 
(v_{k_3 + 2} | w_{2 M - 1 - 2 j_2})_{3/2}
(v_{k_3 + 2} | w_{2 M - 2 j_2})_{3/2}
\, , \\
\mathcal{D} (\bit{u}, \bit{v} | \bit{w}) 
&
=
\prod\limits_{j_1 < k_1 = 2}^{N} 
(u_{j_1} | u_{k_1})_{-1}
\nonumber\\
&\times
\prod\limits_{j_2 < k_2 = 2}^{M} 
(v_{j_2} | v_{k_2})_{-1} (v_{j_2} | v_{k_2})_{-2}
\prod\limits_{j_3 = 1}^{N} \prod\limits_{k_3 = 1}^{M} 
(u_{j_3} | v_{k_3})_{-3/2}
\prod\limits_{j_4 < k_4 = 2}^{N + 2M} 
(w_{j_4} | w_{k_4})_{1}
\, .
\end{align}
This is demonstrated in Fig.\ \ref{MixedMatrixPicture}.

These results are all one needs to extract other singlet pentagon transitions which are allowed by quantum numbers, namely, by sending pairs of (conjugate) hole (fermion and
anifermion) as well as quartets of (anti)fermionic rapidities to infinity. The origin for this limiting procedure is discussed in Sect.\ \ref{NonsingletSection} below.

\section{Moving excitations around}
\label{MovingSection}

To obtain non-diagonal transitions, i.e., involving different number of excitations on the top and the bottom or all of them residing on one side, we have to move particles around 
the contour. For scalars, it is straightforward and is accomplished with the help of the double Wick rotation already used in the derivation of the Mirror equations in Sect.\ 
\ref{MirrorWatsonEqSolutionSection}. We will be interested here in the creation form factor but other cases can be obtained analogously. Starting with the $N$-to-$N$ transition 
\re{HoleHoleTensorPentagon}, every time we get a hole from the bottom to the top side of the pentagon, we acquire one power of $R_{\rm h|h}$. When we move all excitations 
from the bottom to the top, we deduce the form factor in question
\begin{align}
P^{0 | \bit{\scriptstyle I} \bit{\scriptstyle J}} (0 |\bit{u}, \bit{v}) 
=
P^{\bit{\scriptstyle \bar{I}} | \bit{\scriptstyle J}} (\bit{\bar{u}}^{2 \gamma} | \bit{v}) 
= 
\Pi^{0 | \bit{\scriptstyle I} \bit{\scriptstyle J}} (0| \bit{u} , \bit{v})
P_{0|{\rm\bf h}} (0 |\bit{u}, \bit{v})
\, ,
\end{align}
where we used barred notations for reversed order of rapidities $\bit{\bar{u}}  = (u_N, \dots, u_1)$ and O(6) indices $\bit{\bar{I}}  = (I_N, \dots, I_1)$ . Here we stripped the dynamical 
component from the emerging rational prefactors, 
\begin{align}
P_{0|{\rm\bf h}} (0 |\bit{u}, \bit{v})
=
\frac{1
}{
\prod\limits_{i,j}^N P_{\rm h|h} (u_i|v_j) \prod\limits_{i<j}^N P_{\rm h|h} (v_i | v_j) P_{\rm h|h} (u_i | u_j) 
}
\, ,
\end{align}
and shifting them into the SU(4) matrix, which reads as a result
\begin{align}
\label{SingletHoleFF}
\Pi^{0 | \bit{\scriptstyle I} \bit{\scriptstyle J}} (0|\bit{u}, \bit{v})
=
\frac{
\Pi^{\bit{\scriptstyle \bar{I}} | \bit{\scriptstyle J}} (\bit{\bar u} + 2 i | \bit{v})
}{
\prod\limits_{i,j}^N (u_i|v_j)_1 (u_i|v_j)_2
}
\end{align}
in terms of the one determined for the transition amplitude \re{HoleHoleTensor}. For instance, the matrix part of the two-hole creation form factor is
\begin{align}
\Pi^{0| I_1, I_2} (0 | \bit{u}) = \delta^{I_1 I_2} R_{\rm hh} (u_1, u_2)
\, ,
\end{align}
with $ R_{\rm hh} (u_1, u_2)$ given in Eq.\ \re{2GammaHole}.

Fermions, on the other hand, do not have a one-particle mirror transformation. However, from the point of view of the matrix rational prefactor, the modification of changing 
the tensor as one ``moves'' fermions around should not be drastic. We found a useful mnemonic rule, which results in producing a rational factor every time we 
pass the fermion from the initial to the final state
\begin{align}
R_{\Psi \Psi} (u, v)
=
\frac{1}{(u|v)_2}
\, .
\end{align}
Recall that for scalars, the denominator of the rational factor was $(u|v)_1 (u|v)_2$, see Eq.\ \re{2GammaHole}, while for gluons there will be none, i.e., it equals one. The fermion 
is somewhat intermediate between the two and thus was conjectured to have just one factor of particle rapidities. This is analogous to the consideration in Ref.\ \cite{Basso:2014koa} 
where a relation between single-fermion transition and two-fermion form factor was found using similar arguments. The creation form factor of $N$ fermions and $N$ antifermions is
\begin{align}
P^{0 | \bit{\scriptstyle A}}{}_{\bit{\scriptstyle B}} (0 |\bit{u}, \bit{v}) 
=
\Pi^{0 | \bit{\scriptstyle A}}{}_{\bit{\scriptstyle B}} (0| \bit{u} , \bit{v})
P_{0|{\bf \Psi}} (0 |\bit{u}, \bit{v})
\, ,
\end{align}
where the dynamical part, 
\begin{align}
P_{0|{\bf \Psi}} (0 |\bit{u}, \bit{v})
=
\frac{1
}{
\prod\limits_{i,j}^N P_{\rm \Psi | \Psi} (u_i|v_j) \prod\limits_{i<j}^N P_{\rm \Psi | \Psi} (v_i | v_j)  \prod\limits_{i<j}^N P_{\rm \Psi | \Psi} (u_i | u_j) 
}
\, ,
\end{align}
is accompanied by the matrix one
\begin{align}
\Pi^{0 | \bit{\scriptstyle A}}{}_{\bit{\scriptstyle B}} (0|\bit{u} , \bit{v})
=
\frac{
\Pi^{\bit{\scriptstyle \bar{A}} |}{}_{\bit{\scriptstyle B}} (\bit{\bar u} + 2 i | \bit{v})
}{
\prod\limits_{i,j}^N (u_i|v_j)_2
}
\, ,
\end{align}
similar to the rules for the hole excitations.

For the pentagon transitions involving both fermions and holes, one has to add the following mirror transformation
\begin{align}
P_{\rm h|\Psi} (u^{2 \gamma}|v) = \frac{R_{{\rm h}\Psi} (u, v)}{P_{\rm h|\Psi} (u|v)}
\, , \qquad
R_{{\rm h}\Psi} (u, v)
=
\frac{1}{(u|v)_{3/2}}
\, .
\end{align} 
Then the creation form factor of $N$ scalars and $2N$ antifermions is
\begin{align}
P^{0 | \bit{\scriptstyle AB}}{}_{\bit{\scriptstyle C}} (0 |\bit{u}, \bit{v}) 
=
\Pi^{0 | \bit{\scriptstyle AB}}{}_{\bit{\scriptstyle C}} (0| \bit{u} , \bit{v})
P_{0|{\rm\bf h\Psi}} (0 |\bit{u}, \bit{v})
\, ,
\end{align}
where 
\begin{align}
P_{0|{\rm\bf h\Psi}} (0 |\bit{u}, \bit{v})
=
\frac{1
}{
\prod\limits_{i,j}^{N, 2N} P_{\rm h | \Psi} (u_i|v_j) \prod\limits_{i<j}^{2N} P_{\rm \Psi | \Psi} (v_i | v_j) \prod\limits_{i<j}^{N} P_{\rm h | h} (u_i | u_j) 
}
\, ,
\end{align}
and
\begin{align}
\Pi^{0 | \bit{\scriptstyle AB}}{}_{\bit{\scriptstyle C}} (0|\bit{u}, \bit{v})
=
\frac{
\Pi^{\bit{\scriptstyle \bar{A}\bar{B}} |}{}_{\bit{\scriptstyle C}} (\bit{\bar u} + 2 i | \bit{v})
}{
\prod\limits_{i,j}^N (u_i|v_j)_{3/2}
}
\, .
\end{align}

Finally, for the pentagons with all excitations present on its top, using the rules advocated for the fermions and mirror transformation for the holes, we find
\begin{align}
P^{0| \bit{\scriptstyle BCA}}{}_{\bit{\scriptstyle D}} (0| \bit{v}, \bit{u}, \bit{w})
=
\Pi^{0| \bit{\scriptstyle BCA}}{}_{\bit{\scriptstyle D}} (0| \bit{v}, \bit{u}, \bit{w})
P_{0| \rm\bf  h \bar\Psi \Psi} (0 | \bit{v}, \bit{u}, \bit{w})
\, ,
\end{align}
where the dynamical component is
\begin{align}
P_{0| \rm\bf  h \bar\Psi \Psi} (0 | \bit{v}, \bit{u}, \bit{w})
&
=
\frac{
1
}{
\prod\limits_{i<j}^{N + 2M} P_{\Psi | \bar\Psi} (w_i | w_j)
\prod\limits_{i<j}^{N} P_{\Psi | \bar\Psi} (u_i | u_j)
\prod\limits_{i<j}^{M} P_{{\rm h} | {\rm h}} (v_i | v_j)
}
\nonumber\\
&
\times
\frac{
1
}{
\prod\limits_{i,j}^{M,N} P_{{\rm h} | \Psi} (v_i | u_j)
\prod\limits_{i,j}^{N, N + 2M} P_{\Psi | \Psi} (u_i|w_j) 
\prod\limits_{i,j}^{M, N + 2M} P_{{\rm h}|\Psi} (v_i|w_j) 
}
\, ,
\end{align}
while the matrix part reads
\begin{align}
\Pi^{0| \bit{\scriptstyle BCA}}{}_{\bit{\scriptstyle D}} (0| \bit{v}, \bit{u}, \bit{w})
=
\frac{
\Pi^{\bit{\scriptstyle ABC}|}{}_{\bit{\scriptstyle D}} (\bit{\bar{u}} + 2i, \bit{\bar{v}} + 2i | \bit{w})
}{
\prod\limits_{i,j}^{N,N+2M} (u_i|w_j)_{2} \prod\limits_{i,j}^{M,N+2M} (v_i|w_j)_{3/2} 
}
\, .
\end{align}

\section{Nonsinglet pentagons}
\label{NonsingletSection}

Having discussed the singlet pentagons, we are ready to discuss the non-singlet, or charged, transitions. We will not need to address anew equations they obey.
We will provide expressions for these making use of the fact that fermions at zero momentum become supersymmetry generators \cite{Alday:2007mf}. While for the
dynamical component the zero-momentum limit has to be taken on the small fermion sheet \cite{Basso:2010in}, where $p = 0$ corresponds to $u \to \infty$, for the 
rational matrix this can be achieved without performing the analytical continuation and simply taking the infinite-rapidity limit. This allows one to find the ${\bf 4}$ (and 
${\bf\bar{4}}$) pentagons, or rather their tensor structures
\begin{align}
[ \Pi^{A_1}] ^{\dots |}{}_{\dots} (\bit{u}' | \bit{v}) = \lim_{u_1 \to \infty} u_1^{\#} \Pi^{A_1 \dots |}{}_{\dots} (\bit{u} | \bit{v}) 
\, , 
\end{align}
where $\bit{u}'$ is obtained from $\bit{u}$ by removing the rapidity associated with the SU(4) index $A_1$, i.e., $\bit{u}' = \bit{u} \backslash u_1 = (u_2, \dots, u_N)$
and $\#$ is the exponent of the leading power behavior. The latter depends on in- and out-states involved in the transition. In the current study it was empirically 
found on case-by-case basis. It should be possible to derive its generic form for an arbitrary transition, however, we have not succeeded in accomplishing this at
the moment. One can take the limit with respect to any rapidity of fermions involved, yielding the pentagon charged with respect  to the corresponding index. Identical 
considerations give the ${\bf\bar{4}}$-pentagon when one sends corresponding anti-fermion rapidity to infinity or equivalently three fermionic ones.

From the point of view of the matrix structure, two fermions with antisymmetrized indices are identical to the hole insertion.  Therefore, one can extract pentagons charged
with respect to the ${\bf 6}$ of SU(4) by sending corresponding hole rapidity to infinity,
\begin{align}
\label{SextetMatrix}
[\Pi^{I_1}]^{\dots}{}_{\dots} (\bit{u}' | \bit{v}) = \lim_{u_1 \to \infty} u_1^{\#} \Pi^{I_1 \dots}{}_{\dots} (\bit{u} | \bit{v}) 
\, , 
\end{align}
where $\bit{u}'$ is the same as above. Conversion back to the indices in the (anti)fundamental representation can be achieved as discussed in the Footnote \ref{6to6}. Of 
course, in all cases it is irrelevant whether we pick up an excitation sent to infinite rapidity in the initial or final state.

With expressions in hand for \re{FermionicPentagons}, the pentagons involving quartets of anti- and fermions arise from by taking the rapidities of conjugate excitations 
to infinity, e.g., 
\begin{align}
\varepsilon_{B_1 B_2 B_3 B_4} \Pi^{A_1 A_2 A_3 A_4 | 0} (u_1, u_2, u_3, u_4|0) 
= 
\lim_{v \to \infty} v^{\#} \Pi^{A_1 A_2 A_3 A_4 | }{}_{B_1 B_2 B_3 B_4} (u_1, u_2, u_3, u_4 | v, v, v, v)
\, ,
\end{align}
where
\begin{align}
\Pi^{A_1 A_2 A_3 A_4 | 0} (u_1, u_2, u_3, u_4|0) 
=
\frac{\varepsilon^{A_1 A_2 A_3 A_4}}{\prod\limits_{i<j}^4 (u_i | u_j)_{- 1}}
\, .
\end{align}
Another example is demonstrated in Appendix \ref{ExamplesAppendixScalars}.

\section{Gluing up polygons}
\label{IntegrandTestSection}

The known matrix form of pentagon transitions allows one to immediately construct higher polygons. We will address the scalars only as a case of study. It clearly demonstrates
the gluing procedure without the complication of dealing with different flux-tube excitations and their SU(4) indices. The contraction of SU(4) tensors is not a problem for symbolic 
manipulations but the cumbersome form of the output prevents us from displaying final results explicitly in the paper. Thus they are left for the accompanying notebook. 

The contribution of $N_{\rm h}$-hole state propagating in the $N$-gon is
\begin{align}
\label{2Hpolygon}
W_{N}^{N_{\rm h}}
=
\frac{1}{N_{\rm h}!}
\int d \bit{\mu}_{\rm h} 
& 
\
\Pi^{0 | \bit{\scriptstyle I}^{\scriptscriptstyle (1)}} (0 | \bit{u}^{\scriptstyle (1)})
\left[
\prod\limits_{\ell = 1}^{N-6}
\Pi^{\bit{\scriptstyle \bar{I}}{}^{\scriptscriptstyle (\ell)} | \bit{\scriptstyle I}^{\scriptscriptstyle (\ell+1)}}
(- \bit{u}^{\scriptstyle (\ell)} | \bit{u}^{\scriptstyle (\ell+1)} )
\right]
\Pi^{0 | \bit{\scriptstyle \bar{I}}{}^{\scriptscriptstyle (N-5)}} 
(0 | \bit{\bar{u}}^{\scriptstyle (N-5)})
\nonumber\\
\times
\
&
P_{0 | \rm\bf h} (0 | \bit{u}^{\scriptstyle (1)})
\left[
\prod\limits_{\ell = 1}^{N-6}
P_{\rm\bf h|h}
(- \bit{u}^{\scriptstyle (\ell)} | \bit{u}^{\scriptstyle (\ell+1)} )
\right]
P_{0|\rm\bf h} (0 | \bit{\bar{u}}^{\scriptstyle (N-5)})
\, , \nonumber
\end{align}
where the integration measure is conventionally determined by (here for ${\rm p} = {\rm h}$)
\begin{align}
d \bit{\mu}_{\rm h}
=
\prod_{i = 1}^{N_{\rm h}}
\prod_{\ell = 1}^{N-5} d \mu_{\rm h} (u^{\scriptstyle (\ell)}_i)
\, , \qquad
d \mu_{\rm p} (u^{\scriptstyle (\ell)}) = \frac{du}{2 \pi} \mu (u_\ell) {\rm e}^{- \tau_\ell E_{\rm p} (u^{\scriptstyle (\ell)}) + i \sigma_\ell p_{\rm p} (u^{\scriptstyle (\ell)})}
\, ,
\end{align}
with propagation exponents included. Obviously, for the MHV polygon, the creation and annihilation form factors are singlets and have an even number of holes.
While for the NMHV case, the number of scalars is odd since the pentagon operator itself is a sextet of SU(4). The corresponding index gets contracted directly between the top
and bottom as all intermediate pentagons are charge-free. The three-particle case is displayed explicitly in Eq.\ \re{3HsextetFF} of Appendix \ref{ExamplesAppendixScalars}. 
Here and above we used the relation between the creation/annihilation form factors
\begin{align}
P^{\bit{\scriptstyle I} | 0} (\bit{u} | 0) = P^{0 | \bit{\scriptstyle I}} (0 | - \bit{u})
\, .
\end{align}
The contraction of SU(4) matrices can easily be done but results in extremely long expressions due to the factorial growth of the number of functions involved, except, of course, 
for the two-hole case which yields a product of $N-6$ factors $(6 \pi_1 + \pi_2 + \pi_3)$ with arguments as shown in Eq.\ \re{2Hpolygon}. With these results in hand, one can extend 
the consideration of Refs.\ \cite{Basso:2014jfa,Belitsky:2015lzw,Bonini:2016knr} dedicated to the hexagon to the analysis of collinear expansion of any superloop at strong coupling.

To date, there are no explicit results available in the literature for generic number of excitations for polygons with more than six sides. However, we can test our expressions for  
tensor functions against the matrix part of the hexagon proposed in Refs.\ \cite{Basso:2014jfa,Basso:2015uxa}. These consistency cross checks are performed in the accompanying 
notebook for a number of examples with excitations less than ten, the main obstacle for reaching higher numbers being the highly time consuming extraction of the rational function 
from the integral representation in the above papers by taking the residues of the integrand as the number of auxiliary rapidities grows pretty fast.

\section{Conclusions}

In this paper, we developed a constructive method for determination of the internal symmetry group structure of multiparticle (non)singlet pentagons which enter 
as fundamental building blocks in the operator product expansion of scattering amplitudes in maximally supersymmetric Yang-Mills theory. The formalism is based upon analytical
solution of a system of the Mirror and Watson equations obeyed by the corresponding transitions. Their recursive solution reduces all functions accompanying independent tensor 
structures to just one. The latter was conjectured to admit a rational form in terms rapidities of flux-tube excitations and verified a number of tests, which convince us in its correctness.
With this final ingredient in place, the problem of near collinear expansion of scattering amplitudes at any value of the coupling could be viewed as completed. However, it would
nevertheless be highly important to deduce multiple integral representation for contraction of pentagon tensors as they enter the actual scattering amplitudes, generalizing the
earlier consideration for the hexagon \cite{Basso:2015uxa}.

\section*{Acknowledgments}

This research was supported by the U.S. National Science Foundation under the grants PHY-1068286 and PHY-1403891.

\appendix

\section{Explicit examples}
\label{ExamplesAppendix}

Let us provide a few examples for each representative case.

\subsection{Holes}
\label{ExamplesAppendixScalars}

We give the simplest example first, the two-to-two hole transition. The three independent matrix structures are parametrized by two sets of
rapidities $\bit{u} = (u_1, u_2)$ and $\bit{v} = (v_1, v_2)$ for the initial and final states, respectively. The starting point of the recursion is
\begin{align}
\pi_3 (\bit{u} | \bit{v})
=
\frac{(u_1 | v_1)_0 (u_2 | v_2)_1}{(u_1 | u_2)_{-1} (v_1 | v_2)_1}
\, , 
\end{align}
with the other two found from the Watson and Mirror equations
\begin{align}
\pi_2 (\bit{u} | \bit{v})
&
=
\frac{\pi_{3} (\bit{u}|v_2,v_1) 
-
s_{\rm hh}^{(2)} (v_1, v_2)
\pi_{3} (\bit{u}|\bit{v}) 
}{s_{\rm hh}^{(1)} (v_1, v_2)}
\, , \\
\pi_1 (\bit{u} | \bit{v})
&
=
\frac{
(u_1 | v_1)_0
(u_1 | v_1)_{-1}
(u_2 | v_2)_{1}
(u_2 | v_2)_{2}
}{
(u_1 | u_2)_{-1}
(u_1 | u_2)_{-2}
(v_1 | v_2)_{1}
(v_1 | v_2)_{2}
}
\pi_2 (v_1 + 2 i, u_1 | v_2, u_2 + 3 i)
\, ,
\end{align}
respectively. Substituting the explicit expressions for scattering matrix, we find an agreement with the result of Ref.\ \cite{Basso:2013aha} for the case at hand. 

Next, for three-to-three scalar transition parametrized by rapidity arrays $\bit{u} = (u_1, u_2, u_3)$ and $\bit{v} = (v_1, v_2, v_3)$ for the initial and final state, respectively, 
the ``boundary value'' is set by
\begin{align}
\pi_{15} (\bit{u}|\bit{v}) 
=
\frac{
(u_1 | v_1)_0(u_2 | v_1)_0(u_1 | v_2)_0(u_3 | v_2)_1(u_2 | v_3)_1(u_3 | v_3)_1
}{
(u_1 | u_2)_{-1}
(u_1 | u_3)_{-1}
(u_2 | u_3)_{-1}
(v_1 | v_2)_1
(v_1 | v_3)_1
(v_2 | v_3)_1
}
\, .
\end{align}
Then, one can immediately find with the help of a Mathematica routine in the accompanying notebook,
\begin{align}
\pi_{14} (\bit{u}|\bit{v}) 
&=
\frac{\pi_{15} (\bit{u}|v_2,v_1,v_3) 
-
s_{\rm hh}^{(2)} (v_1, v_2)
\pi_{15} (\bit{u}|\bit{v}) 
}{s_{\rm hh}^{(1)} (v_1, v_2)}
\, , \\
\pi_{12} (\bit{u}|\bit{v}) 
&=
\frac{\pi_{15} (\bit{u}|v_1,v_3,v_2) 
-
s_{\rm hh}^{(2)} (v_2, v_3)
\pi_{15} (\bit{u}|\bit{v}) 
}{s_{\rm hh}^{(1)} (v_2, v_3)}
\, , \\
\pi_{11} (\bit{u}|\bit{v}) 
&=
\frac{\pi_{14} (\bit{u}|v_1,v_3,v_2) 
-
s_{\rm hh}^{(2)} (v_2, v_3)
\pi_{14} (\bit{u}|\bit{v}) 
}{s_{\rm hh}^{(1)} (v_2, v_3)}
\, , \\
\pi_{9} (\bit{u}|\bit{v}) 
&=
\frac{\pi_{12} (\bit{u}|v_2,v_1,v_3) 
-
s_{\rm hh}^{(2)} (v_1, v_2)
\pi_{12} (\bit{u}|\bit{v}) 
}{s_{\rm hh}^{(1)} (v_1, v_2)}
\, , \\
\pi_{8} (\bit{u}|\bit{v}) 
&=
\frac{\pi_{11} (\bit{u}|v_2,v_1,v_3) 
-
s_{\rm hh}^{(2)} (v_1, v_2)
\pi_{11} (\bit{u}|\bit{v}) 
}{s_{\rm hh}^{(1)} (v_1, v_2)}
\, , \\
\pi_{5} (\bit{u}|\bit{v}) 
&
=
\frac{
(u_2|u_3)_{-1} (u_2|u_3)_{-2} (u_1|v_1)_0 (u_1|v_1)_{-1} (u_1|v_2)_{0} (u_1|v_2)_{-1} (u_2|v_3)_{1} (u_2|v_3)_{2} (u_3|v_3)_{1} (u_3|v_3)_{2}
}{
(u_1|u_2)_{-1} (u_1|u_2)_{-2} (u_2|u_3)_{-1} (u_2|u_3)_{-2} (u_1|u_3)_{-1} (u_1|u_3)_{-2} (v_1|v_3)_1 (v_1|v_3)_2 (v_2|v_3)_1(v_2|v_3)_2}
\nonumber\\
&
\times
\pi_{14} (v_2 + 2 i, v_1 + 2 i, u_1 | v_3, u_3 + 3 i, u_2 + 3 i) 
\, , \\
\pi_{13} (\bit{u}|\bit{v}) 
&
=
\frac{
(u_1|v_1)_0 (u_1|v_1)_{-1} (u_2|v_1)_0 (u_2|v_1)_{-1} (u_3|v_2)_1 (u_3|v_2)_2 (u_3|v_3)_1 (u_3|v_3)_2
}{
(u_2|u_3)_{-1} (u_2|u_3)_{-2} (u_1|u_3)_{-1} (u_1|u_3)_{-2} (v_1|v_2)_1 (v_1|v_2)_2 (v_1|v_3)_1(v_1|v_3)_2
}
\nonumber\\
&
\times
\pi_{8} (v_1 + 2 i, u_1, u_2| v_2, v_3, u_3 + 3i) 
\, , \\
\pi_{1} (\bit{u}|\bit{v}) 
&
=
\frac{
(u_2|u_3)_{-1} (u_2|u_3)_{-2} (u_1|v_1)_0 (u_1|v_1)_{-1} (u_1|v_2)_{0} (u_1|v_2)_{-1} (u_2|v_3)_{1} (u_2|v_3)_{2} (u_3|v_3)_{1} (u_3|v_3)_{2}
}{
(u_1|u_2)_{-1} (u_1|u_2)_{-2} (u_2|u_3)_{-1} (u_2|u_3)_{-2} (u_1|u_3)_{-1} (u_1|u_3)_{-2} (v_1|v_3)_1 (v_1|v_3)_2 (v_2|v_3)_1(v_2|v_3)_2}
\nonumber\\
&
\times
\pi_{8} (v_2 + 2 i, v_1 + 2 i, u_1 | v_3, u_3 + 3 i, u_2 + 3 i) 
\, , \\
\pi_{10} (\bit{u}|\bit{v}) 
&
=
\frac{
(u_1|v_1)_0 (u_1|v_1)_{-1} (u_2|v_1)_0 (u_2|v_1)_{-1} (u_3|v_2)_1 (u_3|v_2)_2 (u_3|v_3)_1 (u_3|v_3)_2
}{
(u_2|u_3)_{-1} (u_2|u_3)_{-2} (u_1|u_3)_{-1} (u_1|u_3)_{-2} (v_1|v_2)_1 (v_1|v_2)_2 (v_1|v_3)_1(v_1|v_3)_2
}
\nonumber\\
&
\times
\pi_{11} (v_1 + 2 i, u_1, u_2| v_2, v_3, u_3 + 3i) 
\, , \\
\pi_{2} (\bit{u}|\bit{v}) 
&
=
\frac{
(u_2|u_3)_{-1} (u_2|u_3)_{-2} (u_1|v_1)_0 (u_1|v_1)_{-1} (u_1|v_2)_{0} (u_1|v_2)_{-1} (u_2|v_3)_{1} (u_2|v_3)_{2} (u_3|v_3)_{1} (u_3|v_3)_{2}
}{
(u_1|u_2)_{-1} (u_1|u_2)_{-2} (u_2|u_3)_{-1} (u_2|u_3)_{-2} (u_1|u_3)_{-1} (u_1|u_3)_{-2} (v_1|v_3)_1 (v_1|v_3)_2 (v_2|v_3)_1(v_2|v_3)_2}
\nonumber\\
&
\times
\pi_{11} (v_2 + 2 i, v_1 + 2 i, u_1 | v_3, u_3 + 3 i, u_2 + 3 i) 
\, , \\
\pi_{6} (\bit{u}|\bit{v}) 
&
=
\frac{
(u_1|v_1)_0 (u_1|v_1)_{-1} (u_2|v_1)_0 (u_2|v_1)_{-1} (u_3|v_2)_1 (u_3|v_2)_2 (u_3|v_3)_1 (u_3|v_3)_2
}{
(u_2|u_3)_{-1} (u_2|u_3)_{-2} (u_1|u_3)_{-1} (u_1|u_3)_{-2} (v_1|v_2)_1 (v_1|v_2)_2 (v_1|v_3)_1(v_1|v_3)_2
}
\nonumber\\
&
\times
\pi_{9} (v_1 + 2 i, u_1, u_2| v_2, v_3, u_3 + 3i) 
\, , \\
\pi_{4} (\bit{u}|\bit{v}) 
&
=
\frac{
(u_2|u_3)_{-1} (u_2|u_3)_{-2} (u_1|v_1)_0 (u_1|v_1)_{-1} (u_1|v_2)_{0} (u_1|v_2)_{-1} (u_2|v_3)_{1} (u_2|v_3)_{2} (u_3|v_3)_{1} (u_3|v_3)_{2}
}{
(u_1|u_2)_{-1} (u_1|u_2)_{-2} (u_2|u_3)_{-1} (u_2|u_3)_{-2} (u_1|u_3)_{-1} (u_1|u_3)_{-2} (v_1|v_3)_1 (v_1|v_3)_2 (v_2|v_3)_1(v_2|v_3)_2}
\nonumber\\
&
\times
\pi_{9} (v_2 + 2 i, v_1 + 2 i, u_1 | v_3, u_3 + 3 i, u_2 + 3 i) 
\, , \\
\pi_{7} (\bit{u}|\bit{v}) 
&=
\frac{\pi_{10} (\bit{u}|v_2,v_1,v_3) 
-
s_{\rm hh}^{(2)} (v_1, v_2)
\pi_{10} (\bit{u}|\bit{v}) 
}{s_{\rm hh}^{(1)} (v_1, v_2)}
\, , \\
\pi_{3} (\bit{u}|\bit{v}) 
&=
\frac{\pi_{2} (\bit{u}|v_1,v_3,v_2) 
-
s_{\rm hh}^{(2)} (v_2, v_3)
\pi_{2} (\bit{u}|\bit{v}) 
}{s_{\rm hh}^{(1)} (v_2, v_3)}
\, .
\end{align}

In Sect.\ \ref{IntegrandTestSection} dedicated to the construction of higher polygons, we need an expression for the charged pentagon creation form factor. As we explained in 
Sect.\ \ref{MovingSection}, all nonsinglet pentagons can be found from the singlet ones. For the case at hand, we use the $2 \to 2$ hole transition and move all excitations 
to the top according to Eq.\ \re{SextetMatrix} and then, sending one of the rapidities there to infinity, we find, making use of the result \re{SingletHoleFF},
\begin{align}
\label{3HsextetFF}
[ \Pi^I]^{0 | I_1 I_2 I_3} (0 | \bit{v}) 
=
\delta^{I_1 I_2} \delta^{I_3 I} R^{(1)}_{\rm hhh} (\bit{v}) 
+
\delta^{I_2 I_3} \delta^{I_1 I} R^{(2)}_{\rm hhh} (\bit{v}) 
+
\delta^{I_1 I_3} \delta^{I_2 I} R^{(3)}_{\rm hhh} (\bit{v}) 
\, .
\end{align}
Here
\begin{align}
R^{(1)}_{\rm hhh} (\bit{v}) 
&= \frac{(v_1|v_3)_{3}}{(v_1|v_2)_{1} (v_1|v_2)_{2} (v_1|v_3)_{1} (v_1|v_3)_{2}(v_2|v_3)_{1}}
\, , \\
R^{(2)}_{\rm hhh} (\bit{v}) 
&= \frac{(v_1|v_3)_{- 3}}{(v_1|v_2)_{1} (v_2|v_3)_{1} (v_2|v_3)_{2} (v_1|v_3)_{1} (v_1|v_3)_{2}}
\, , \\
R^{(3)}_{\rm hhh} (\bit{v}) 
&=
- \frac{1}{(v_1|v_2)_{1} (v_1|v_3)_{1} (v_1|v_3)_{2} (v_2|v_3)_{1}}
\, .
\end{align}
Pentagons and form factors with larger number of particles are found in a similar manner.

\subsection{Fermions}
\label{ExamplesAppendixFermions}

For the transition of three fermions with rapidities $\bit{u} = (u_1, u_2, u_3)$ to three antifermions with $\bit{v} = (v_1, v_2, v_3)$, we have
\begin{align}
\pi_6 (\bit{u} | \bit{v})
=
\frac{
(u_1|v_1)_0 (u_2|v_1)_0 (u_1|v_2)_0 (u_3|v_2)_1 (u_2|v_3)_1 (u_3|v_3)_1
}{
(u_1|u_2)_{-1} (u_1|u_3)_{-1} (u_2|u_3)_{-1} (v_1|v_2)_1 (v_1|v_3)_1 (v_2|v_3)_1
}
\, ,
\end{align}
and the rest are found from Watson equations
\begin{align}
\pi_5 (\bit{u} | \bit{v})
&
=
\frac{
\pi_6 (\bit{u} | v_2,v_1,v_3)
-
s_{\Psi\Psi}^{(2)} (v_1, v_2)
\pi_6 (\bit{u}|\bit{v}) 
}{s_{\Psi\Psi}^{(1)} (v_1, v_2)}
\, , \\
\pi_4 (\bit{u} | \bit{v})
&
=
\frac{\pi_6 (\bit{u} | v_1,v_3,v_2)
-
s_{\Psi\Psi}^{(2)} (v_2, v_3)
\pi_6 (\bit{u}|\bit{v}) 
}{s_{\Psi\Psi}^{(1)} (v_2, v_3)}
\, , \\
\pi_3 (\bit{u} | \bit{v})
&
=
\frac{\pi_5 (\bit{u} | v_1,v_3,v_2)
-
s_{\Psi\Psi}^{(2)} (v_2, v_3)
\pi_5 (\bit{u}|\bit{v}) 
}{s_{\Psi\Psi}^{(1)} (v_2, v_3)}
\, , \\
\pi_2 (\bit{u} | \bit{v})
&
=
\frac{
\pi_4 (\bit{u} | v_2,v_1,v_3)
-
s_{\Psi\Psi}^{(2)} (v_1, v_2)
\pi_4 (\bit{u}|\bit{v}) 
}{s_{\Psi\Psi}^{(1)} (v_1, v_2)}
\, , \\
\pi_1 (\bit{u} | \bit{v})
&
=
\frac{
\pi_3 (\bit{u} | v_2,v_1,v_3)
-
s_{\Psi\Psi}^{(2)} (v_1, v_2)
\pi_3 (\bit{u}|\bit{v}) 
}{s_{\Psi\Psi}^{(1)} (v_1, v_2)}
\, .
\end{align}

\subsection{Holes and (anti)fermions}
\label{ExamplesAppendixHolesFermions}

We move on to the final two examples. To start with, let us present expressions for two holes $\bit{u} = (u_1, u_2)$ to four fermions $\bit{v} = (v_1, v_2, v_3, v_4)$ transitions. 
The seed for the recursion is
\begin{align}
\pi_6 (\bit{u} | \bit{v})
&=
\frac{(u_1|v_1)_{-1/2} (u_1|v_2)_{-1/2} (u_2|v_3)_{3/2} (u_2|v_4)_{3/2}}{(u_1|u_2)_{-1} (u_1|u_2)_2 (v_1|v_2)_1 (v_1|v_3)_1 (v_1|v_4)_1 (v_2|v_3)_1 (v_2|v_4)_1 (v_3|v_4)_1}
\, ,
\end{align}
with the rest being
\begin{align}
\pi_5 (\bit{u} | \bit{v})
&
=
\frac{\pi_6 (\bit{u} | v_1, v_3, v_2, v_4)
-
s_{\Psi\Psi}^{(2)} (v_2, v_3)
\pi_6 (\bit{u}|\bit{v}) 
}{s_{\Psi\Psi}^{(1)} (v_2, v_3)}
\, , \\
\pi_3 (\bit{u} | \bit{v})
&
=
\frac{
\pi_5 (\bit{u} | v_1,v_3, v_2,v_4)
-
s_{\Psi\Psi}^{(2)} (v_1, v_2)
\pi_5 (\bit{u}|\bit{v}) 
}{s_{\Psi\Psi}^{(1)} (v_1, v_2)}
\, , \\
\pi_4 (\bit{u} | \bit{v})
&
=
\frac{
\pi_5 (\bit{u} | v_1, v_2, v_4, v_3)
-
s_{\Psi\Psi}^{(2)} (v_3, v_4)
\pi_5 (\bit{u}|\bit{v}) 
}{s_{\Psi\Psi}^{(1)} (v_3, v_4)}
\, , \\
\pi_2 (\bit{u} | \bit{v})
&
=
\frac{
\pi_4 (\bit{u} | v_2, v_1, v_3, v_4)
-
s_{\Psi\Psi}^{(2)} (v_1, v_2)
\pi_4 (\bit{u}|\bit{v}) 
}{s_{\Psi\Psi}^{(1)} (v_1, v_2)}
\, , \\
\pi_1 (\bit{u} | \bit{v})
&
=
\frac{
\pi_2 (\bit{u} | v_1,v_3,v_2, v_4)
-
s_{\Psi\Psi}^{(2)} (v_2, v_3)
\pi_2 (\bit{u}|\bit{v}) 
}{s_{\Psi\Psi}^{(1)} (v_2, v_3)}
\, .
\end{align}

Now, we demonstrate the case of two fermions and a hole on the bottom along with four antifermions on the top. The twisted component is
\begin{align}
\pi_{12} (\bit{u}, \bit{v} | \bit{w})
=
\frac{
(u_1|w_1)_0 (u_1|w_2)_0 (u_1|w_3)_0 (u_2|w_1)_0 (u_2|w_2)_0 (u_2|w_4)_1 (v_1|w_3)_{3/2} (v_1|w_4)_{3/2}
}{
(u_1|u_2)_{-1} (u_1|v_1)_{-3/2} (u_2|v_1)_{- 3/2} (w_1|w_2)_{1} (w_1|w_3)_{1} (w_1|w_4)_{1} (w_2|w_3)_{1} (w_2|w_4)_{1} (w_3|w_4)_{1}
}
\, ,
\end{align}
with the remaining ones emerging from the Watson equations,
\begin{align}
\pi_{11} (\bit{u}, \bit{v} | \bit{w})
&
=
\frac{\pi_{12} (\bit{u} , \bit{v} | w_1, w_3, w_2, w_4 )
-
s_{\Psi\Psi}^{(2)} (w_2, w_3)
\pi_{12} (\bit{u}, \bit{v} | \bit{w}) 
}{s_{\Psi\Psi}^{(1)} (w_2, w_3)}
\, , \\
\pi_{9} (\bit{u}, \bit{v} | \bit{w})
&
=
\frac{\pi_{12} (\bit{u} , \bit{v} | w_1, w_2, w_4, w_3 )
-
s_{\Psi\Psi}^{(2)} (w_3, w_4)
\pi_{12} (\bit{u}, \bit{v} | \bit{w}) 
}{s_{\Psi\Psi}^{(1)} (w_3, w_4)}
\, , \\
\pi_{10} (\bit{u}, \bit{v} | \bit{w})
&
=
\frac{\pi_{11} (\bit{u} , \bit{v} | w_2, w_1, w_3, w_4 )
-
s_{\Psi\Psi}^{(2)} (w_1, w_2)
\pi_{11} (\bit{u}, \bit{v} | \bit{w}) 
}{s_{\Psi\Psi}^{(1)} (w_1, w_2)}
\, , \\
\pi_{8} (\bit{u}, \bit{v} | \bit{w})
&
=
\frac{\pi_{11} (\bit{u} , \bit{v} | w_1, w_2, w_4, w_3)
-
s_{\Psi\Psi}^{(2)} (w_3, w_4)
\pi_{11} (\bit{u}, \bit{v} | \bit{w}) 
}{s_{\Psi\Psi}^{(1)} (w_3, w_4)}
\, , \\
\pi_{7} (\bit{u}, \bit{v} | \bit{w})
&
=
\frac{\pi_{10} (\bit{u} , \bit{v} | w_1, w_2, w_4, w_3)
-
s_{\Psi\Psi}^{(2)} (w_3, w_4)
\pi_{10} (\bit{u}, \bit{v} | \bit{w}) 
}{s_{\Psi\Psi}^{(1)} (w_3, w_4)}
\, , \\
\pi_{6} (\bit{u}, \bit{v} | \bit{w})
&
=
\frac{\pi_{9} (\bit{u} , \bit{v} | w_1, w_3, w_2, w_4)
-
s_{\Psi\Psi}^{(2)} (w_2, w_3)
\pi_{9} (\bit{u}, \bit{v} | \bit{w}) 
}{s_{\Psi\Psi}^{(1)} (w_2, w_3)}
\, , \\
\pi_{5} (\bit{u}, \bit{v} | \bit{w})
&
=
\frac{\pi_{8} (\bit{u} , \bit{v} | w_1, w_3, w_2, w_4)
-
s_{\Psi\Psi}^{(2)} (w_2, w_3)
\pi_{8} (\bit{u}, \bit{v} | \bit{w}) 
}{s_{\Psi\Psi}^{(1)} (w_2, w_3)}
\, , \\
\pi_{4} (\bit{u}, \bit{v} | \bit{w})
&
=
\frac{\pi_{7} (\bit{u} , \bit{v} | w_1, w_3, w_2, w_4)
-
s_{\Psi\Psi}^{(2)} (w_2, w_3)
\pi_{7} (\bit{u}, \bit{v} | \bit{w}) 
}{s_{\Psi\Psi}^{(1)} (w_2, w_3)}
\, , \\
\pi_{3} (\bit{u}, \bit{v} | \bit{w})
&
=
\frac{\pi_{6} (\bit{u} , \bit{v} | w_2, w_1, w_3, w_4)
-
s_{\Psi\Psi}^{(2)} (w_1, w_2)
\pi_{6} (\bit{u}, \bit{v} | \bit{w}) 
}{s_{\Psi\Psi}^{(1)} (w_1, w_2)}
\, , \\
\pi_{2} (\bit{u}, \bit{v} | \bit{w})
&
=
\frac{\pi_{5} (\bit{u} , \bit{v} | w_2, w_1, w_3, w_4)
-
s_{\Psi\Psi}^{(2)} (w_1, w_2)
\pi_{5} (\bit{u}, \bit{v} | \bit{w}) 
}{s_{\Psi\Psi}^{(1)} (w_1, w_2)}
\, , \\
\pi_{1} (\bit{u}, \bit{v} | \bit{w})
&
=
\frac{\pi_{4} (\bit{u} , \bit{v} | w_2, w_1, w_3, w_4)
-
s_{\Psi\Psi}^{(2)} (w_1, w_2)
\pi_{4} (\bit{u}, \bit{v} | \bit{w}) 
}{s_{\Psi\Psi}^{(1)} (w_1, w_2)}
\, .
\end{align}

 %%%%%%  Bibliography %%%%%%%%%%%%%%%%%%%%%%%%%%%%%%%%%%%%%%%%%%%%

%%%%%%%%%%%%%%%%%%%%%%%%%%%%%%%%%%%%%%%%%%%%%%%%%%%%%%%%%%%%%%%%
\end{document}